\documentclass[aps,pra,twocolumn,showpacs,showkeys,footinbib,superscriptaddress]{revtex4-1}

\usepackage{amsmath}
\usepackage{amssymb}
\usepackage{graphicx}
\usepackage{bm}
\usepackage{color}
\usepackage{relsize}
\usepackage{braket}
\usepackage[caption=false]{subfig}

\usepackage{hyperref}
\usepackage[all]{hypcap}

\begin{document}
\title{Turning many-body problems to few-body ones in photoexcited semiconductors using the stochastic variational method in momentum space, SVM-{\it k} }
\date{\today}

\author{Dinh Van Tuan}
\email[]{vdinh@ur.rochester.edu}
\affiliation{Department of Electrical and Computer Engineering, University of Rochester, Rochester, New York 14627, USA}
\author{Hanan~Dery}
\affiliation{Department of Electrical and Computer Engineering, University of Rochester, Rochester, New York 14627, USA}
\affiliation{Department of Physics and Astronomy, University of Rochester, Rochester, New York 14627, USA}

\begin{abstract} 
We develop an efficient computational technique to calculate composite excitonic states in photoexcited semiconductors through the stochastic variational method (SVM). Many-body interactions between an electron gas and the excitonic state are embodied in the problem through Fermi holes in the conduction band, introduced when electrons are pulled out of the Fermi sea to bind the photoexcited electron-hole pair. We consider the direct Coulomb interaction between distinguishable particles in the complex, the exchange-induced band-gap renormalization effect, and electron-hole exchange interaction between an electron and its conduction-band hole. We provide analytical expressions for potential matrix elements, using a technique that allows us to circumvent the difficulty imposed by the occupation of low-energy electron states in the conduction band. We discuss the computational steps one should implement in order to perform the calculation, and how to extract kinetic energies of individual particles in the complex, average inter-particle distances, and density distributions.
\end{abstract}

\maketitle

\section{Introduction}
The variational method is a common technique to solve the Schr\"{o}dinger Equation of few-body systems \cite{McMillan_PRA65,Ceperley_RMP95, Foulkes_RMP01, Mitroy_RMP13,SchererBook}. This method can be used to study excitonic states such as the neutral exciton or biexciton in photoexcited semiconductors \cite{Riva_PRB2000,Berkelbach_PRB13,Mayers_PRB15,Kidd_PRB16,Donck_PRB17,Mostaani_PRB17,VanTuan_PRB18}. In the limit that the semiconductor has one electron in the conduction band (CB) prior to photoexcitation, the method can be used to study negative trions, wherein two CB electrons bind to a valence band (VB) hole. Equivalently, positive trions can be studied  when two VB holes bind to an electron in the CB. We will continue the discussion by considering electron-doped semiconductors, bearing in mind that equivalent discussion can be drawn for hole-doped semiconductors. 

Rather than having one electron in the CB, practical settings include an interacting electron gas that occupies the low energy states of the CB. Thus, we face a true many-body problem for which solving the Schr\"{o}dinger Equation  is hopeless.  Yet, experiments show that many-body signatures evolve from the trion optical transition at small electron densities \cite{Finkelstein_PRB96, Andronikov_PRB05,Koudinov_PRL14,Wang_NanoLett17,Smolenski_PRL19,Liu_NatComm21,Liu_PRL20,Wang_PRX20,Li_NanoLett22}, indicating that the trion is a good starting point for theoretical analysis. One can then study the interaction between trions and Fermi-sea electrons through the optical  susceptibility function \cite{Bronold_PRB00,Suris_PSS01,Esser_pssb01}, where the outcome is a correlated trion state, which can also be studied by variational methods \cite{Chang_PRB18,Rana_PRB20}. The correlated trion state is a four-body composite (Suris tetron), in which the bare trion is bound to a Fermi hole. Namely,  the trion and the lack of Fermi-sea electrons in its vicinity move together. The theory we present in this paper extends this concept further.

The creation of trions in electron-doped semiconductors comes from the presence of the VB hole, without which two electrons would keep apart. CB electrons can bind the VB hole if they can scatter between unoccupied $k$-states, enabling them to orbit and stay close to the VB hole. To do so, the electron has to vacate a state below the Fermi surface and sample a relatively large portion of the $k$-space above it. Just as promoting an electron from the VB to the CB during photoexcitation leaves behind an unfilled state in the VB (hole), pulling out an electron from below the Fermi surface leaves behind a hole in the Fermi sea. This CB hole stays close to the pulled out electron by scattering with electrons in occupied $k$-states below the Fermi surface. 

\begin{figure}
\centering
\includegraphics*[width=7cm]{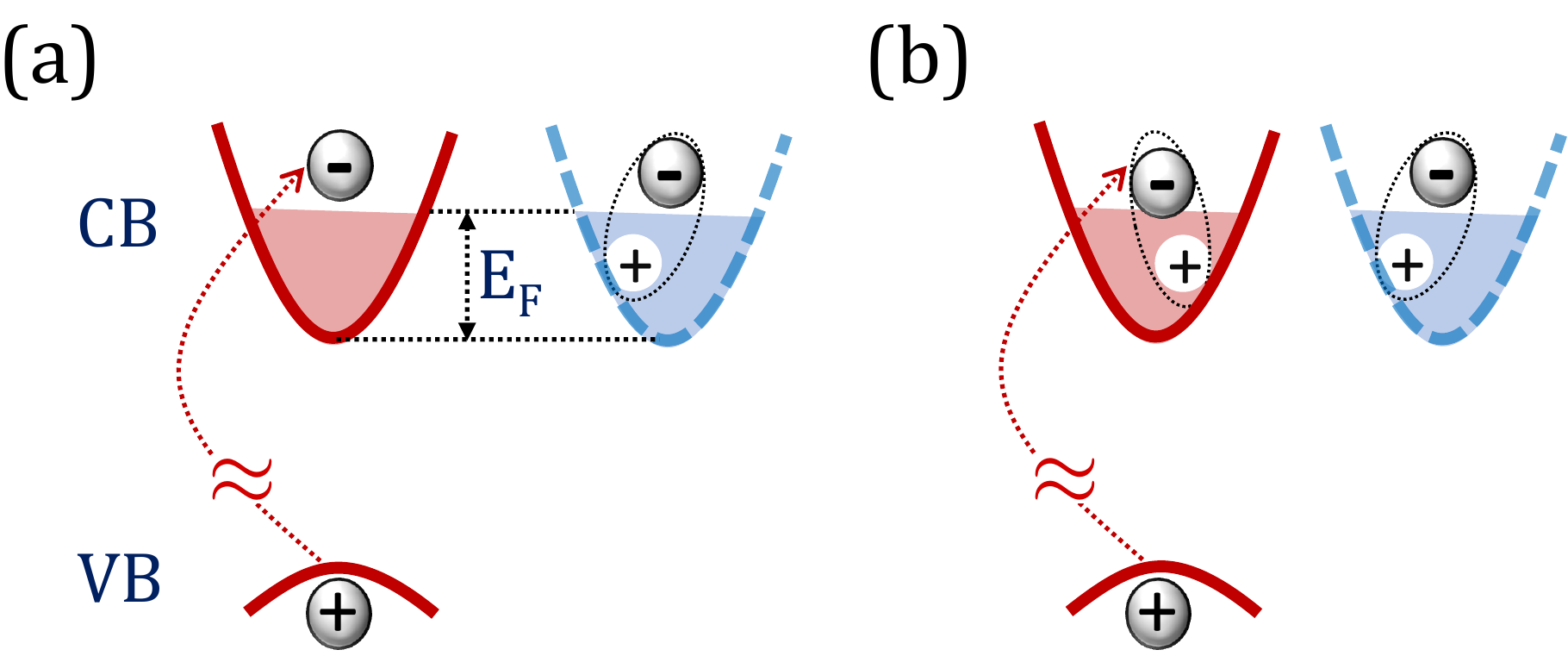}
\caption{Correlated trions in photoexcited semiconductors with (a)/(b) one/two CB holes.}\label{fig:scheme} 
\end{figure}

Figure~\ref{fig:scheme} shows two configurations for correlated trions in electron-doped semiconductor. The left diagram shows the tetron, wherein the photoexcited pair binds to an  electron-hole CB pair.  The photoexcited electron interacts with similar-spin electrons from the same valley through exchange interaction, leading to band-gap renormalization (BGR). The interaction lowers the energy of electrons with similar quantum numbers by keeping them further apart. The CB electron in the other valley is accompanied by CB hole, which serves the same function. Namely, it is an expression for the lack of electrons from the other valley around the complex. Overall, the tetron restores charge neutrality, consistent with the fact that photoexcitation neither adds nor removes charge from the semiconductor.   

Figure~\ref{fig:scheme}(b) shows a 5-body configuration of the correlated trion, wherein two CB holes accompany the trion. The CB hole in the valley of the photoexcited electron can be created by  a shakeup process, during which electrons with the same  spin and valley of the photoexcited electron are driven away. The BGR effect is weak in the five-particle complex because the exchange interaction of an electron above the Fermi level is largely offset by that of the missing electron below the Fermi level (i.e., of the CB hole). In the five-particle complex, the binding of the second CB hole to the trion replaces the functionality of BGR in lowering the total energy. While it is not clear at this point which of the two configurations of Fig.~\ref{fig:scheme} better reflects the underlying physics, the energy is lowered in both cases.

In multi-valley  semiconductors, the concept of composite excitonic states can be extended to complexes with more than two electron-hole pairs in the CB.  Each electron in the complex  comes with distinct valley and spin quantum numbers, allowing the electrons to stay together near the VB hole at the same time without violating the Pauli exclusion principle. In this work, we present an efficient computational technique to calculate these states using the stochastic variational method in momentum space (SVM-$k$). The SVM was originally developed by Varga and Suzuki to study the Schr\"{o}dinger Equation of few-body systems \cite{Varga_NuPhys1994,VargaBook,Varga1997,Varga2008}. The many-body problem in our case is recast to a problem with few quasiparticles, where interactions between Fermi-sea electrons and the excitonic state are embodied through CB holes. In addition to BGR and direct Coulomb interactions between distinguishable particles in the complex, we account for the electron-hole exchange interaction between an electron and its CB hole.

The organization of the paper is as follows. We delineate the details of the SVM-$k$ model in Sec.~\ref{sec:exp}, focusing on composite excitonic states in electron-doped semiconductors. Using second quantization, we first present the basis states and the Gaussian envelope functions associated with these states, followed by calculation of the resulting kinetic and potential matrix elements. To make the computational complexity tractable, the calculation of Coulomb interaction integrals is performed analytically using a technique that allows us to circumvent the difficulty imposed by the occupied low-energy states in the CB. We show analytical results using the Keldysh-Rytove potential in two-dimensional systems and explain the procedure one should take in cases of other potential forms. Finally, we discuss the computational steps of the variational method that one should implement in order to perform the calculation. Section~\ref{sec:DisCon} includes outlook and conclusions. Interested readers can find applications  of the SVM-$k$ model in Refs.~\cite{s,g,h}, where we study 4, 5 and 6-body excitonic complexes in monolayer transition-metal dichalcogenides. Here, the focus is on theoretical formulation and computational aspects.   

\section{The SVM-\lowercase{\textit{k}} model}\label{sec:exp}

To take  the filling factor of the Fermi sea into account, we use second quantization and write the Hamiltonian in momentum space ($\hbar =1$)
\begin{eqnarray}
H&=&K+V=\sum_{{\bf k}_\alpha} \frac{k^2}{2 m_\alpha} c^\dagger_{{\bf k}_\alpha} c_{{\bf k}_\alpha} \nonumber \\
&+&\frac{1}{2} \sum_{{\bf k}_\alpha, {\bf p}_\beta,{\bf q}} V_{\alpha,\beta}({\bf q}) c^\dagger_{{\bf k}_\alpha + \bf{q}} c^\dagger_{\bf{p}_\beta - \bf{q}} c_{{\bf p}_\beta} c_{{\bf k}_\alpha} \,.\label{eq:H}
\end{eqnarray}
$c^\dagger_{{\bf k}_\alpha}  $ ($c_{{\bf k}_\alpha}$) is the creation (annihilation) operator of an electron with momentum $\bf k$, and the index $\alpha$ encompasses the band index, spin, and valley quantum numbers. $V_{\alpha,\beta}({\bf q})$ is the Coulomb potential. 

To study how excitonic states emerge from the general Hamiltonian in Eq.~(\ref{eq:H}), we will consider an excitonic complex made of $2(N+1)$ quasiparticles. Two of the quasiparticles come from the photoexcited electron-hole pair when light with momentum $\bf Q$ promotes an electron from the VB to CB, leaving behind a VB hole. The photoexcitation is accompanied by excitation of $N$ other electron-hole pairs in the CB Fermi sea. Each of the $N+1$ electrons in the complex comes with distinct quantum numbers.

\subsection{Basis states of the quasiparticle system}

The eigenstate of the excitonic complex is written as a linear superposition 
\begin{equation}
| \psi \rangle = \sum_{i} C_i | \phi_i \rangle,
\label{Eq:WaveF}
\end{equation}
where the basis states are
\begin{equation}
| \phi_i \rangle = \sum_{\bf X} \phi_i({\bf X}) \,\,\,  c^\dagger_{{\bf k}_0} c_{v,{\bf p}_0} \prod_{\ell=1}^{N}  c^\dagger_{{\bf k}_\ell}  c_{{\bf p}_\ell} |  \varphi_0 \rangle.
\label{Eq:TrialWave-k}
\end{equation}
$|\varphi_0 \rangle$ is the ground state of the system before light excitation with filled electronic states up to  $E_F$.  ${\bf X} = \{ {\bf k}_0, {\bf k}_1, -{\bf p}_1, ..., {\bf k}_{N}, -{\bf p}_{N} \}$ is a set of $(2N+1)$ momenta variables of the photoexcited electron ($\mathbf{k}_0$) and other electron-hole pairs ($\mathbf{k}_\ell,-\mathbf{p}_\ell$). The momentum of the VB hole, $-{\bf p}_0$, is extracted from $\sum_{\ell=0}^{N} \left( {\bf k}_\ell - {\bf p}_\ell \right)= {\bf Q}$, where $\bf Q$ is the center-of-mass (CoM) momentum transferred to the system from light absorption.  In what follows, we assume $Q =0$  due to the minute photon momentum. Furthermore, since $\bf{Q}$ is a constant of motion, ${\bf X}$ includes $(2N+1)$ rather than $2(N+1)$  components, meaning that  the Schr\"{o}dinger Equation we will solve includes $(2N+1)$ degrees of freedom. We will use the notation $\ell$ and $\bar{\ell}$ to represent matrix indices associated with the $\ell^\text{th}$ electron and its CB hole, respectively. Accordingly, ${\bf X}_\ell = {\bf k}_\ell$ and ${\bf X}_{\bar{\ell}} = -{\bf p}_\ell$.  Lastly, the basis states in Eq.~(\ref{Eq:TrialWave-k})  include correlated Gaussian functions \cite{Mitroy_RMP13}, which in  momentum space read
\begin{equation}
\phi_i({\bf X}) = \exp \left( -\frac{1}{2} {\bf X}^\text{T} M_i {\bf X}\right).
\label{Eq:GaussBas}
\end{equation} 
$M_i$ is a $(2N+1) \times (2N+1)$ symmetric, real, and positive definite matrix. Off-diagonal matrix elements of $M_i$ represent correlations between momenta of two corresponding quasiparticles, where neither is the VB hole. As the latter is not part of ${\bf X}$,  its physical parameters are dealt with differently. This point will become clear later. 

Assuming $\mathcal{N}_\text{b}$ basis states,  the ground-state of the $2(N+1)$-quasiparticle system is obtained from solution of the matrix equation
\begin{equation}
H C = E O C\,,
\label{Eq:MatrixEq}
\end{equation}
where $C = \{ C_1, ...,C_{\mathcal{N}_\text{b}} \}^\text{T}$ is a column vector of the coefficients in Eq.(\ref{Eq:WaveF}).  We find these coefficients and the energy of the system by treating all elements of the matrices $M_i$ as variational parameters, to be found through the energy minimization process of the $2(N+1)$-quasiparticle complex.  $O$ is the overlap matrix with elements
\begin{equation}
O_{ij} =  \langle \phi_i | \phi_j \rangle =  \sum_{\bf X} \phi^*_i({\bf X}) \phi_j({\bf X}) F\left( {\bf X} \right).
\end{equation}
The filling factor $F\left( {\bf X} \right)$ handles the momentum space restriction for all quasiparticles in the system
\begin{equation}
F\left( {\bf X} \right) = \left( 1 - f_{{\bf k}_0}\right) f_{{\bf p}_0} \prod_{\ell=1}^{N} \left( 1 - f_{{\bf k}_\ell} \right) f_{{\bf p}_\ell}.
\end{equation}
At zero temperature, electrons (holes) of the complex are limited to $k_\ell > k_{F,\ell}$ ($p_\ell < k_{F,\ell}$), where $k_{F,\ell}$ is the Fermi wavenumber of the energy pocket in which the $\ell^{\text{th}}$ electron (hole) resides.

\begin{figure}
\centering
\includegraphics*[width=8cm]{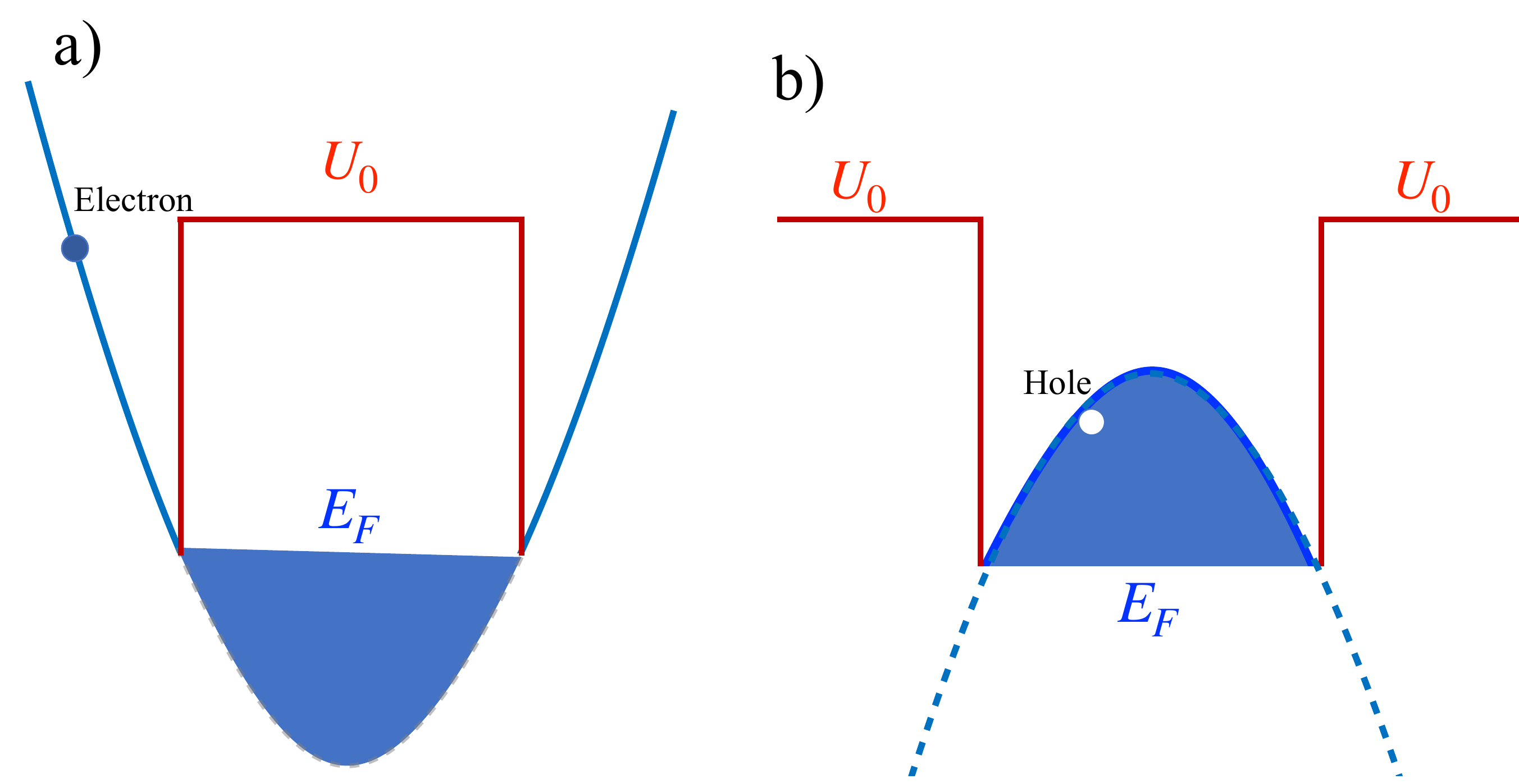}
\caption{Band structure modifications for conduction-band (a) electrons and (b) holes. This modification is introduced in lieu of using the filling factor. }\label{fig:Fig1} 
\end{figure}

The phase space of $F\left( {\bf X} \right)$ renders analytical calculations of energy matrix elements impossible. {\it To overcome this difficulty, we modify the band structures such that the kinetic energy of an electron (hole) below (above) the Fermi level is large}. Figures~\ref{fig:Fig1}(a) and (b) show the modified band structure for electrons and Fermi holes, respectively, using step functions with energy $U_0$ below (above) the Fermi levels for electrons (holes).  By choosing large $U_0$ compared with the energy of the $2(N+1)$-quasiparticle complex, we eliminate solutions in which the $\ell^{\text{th}}$ electron penetrates the prohibited region $k_\ell < k_{F,\ell}$ and its hole penetrates the complementary prohibited region, $p_\ell> k_{F,\ell}$. That is, the energy minimization process is forced to choose solutions in which electrons (holes) of the complex are kept above (below) the Fermi energy.  In addition to the kinetic and potential energies in Eq.~(\ref{eq:H}), the band-structure modification corresponds to additional potentials for electron $\ell$ and its CB hole $\bar{\ell}$,
\begin{eqnarray}
U_\ell({\bf X}_\ell) &=& \left(U_0 - \frac{k_\ell^2}{2 m_\ell} \right) \Theta \left(  k_F  -  k_\ell \right), \nonumber \\
U_{\bar{\ell}}({\bf X}_{\bar{\ell}})  &=& \left(U_0 + \frac{p_{\ell}^2}{2 m_\ell} \right) \Theta \left( p_{\ell} -k_F \right). 
\label{Eq:ArtiPoten}
\end{eqnarray}

\subsection{Matrix elements between basis states} \label{sec:matrix_elements}
With the help of the band structure modification we can set $F({\bf X}) =  1$ in all formulas and perform analytical calculations for all matrix elements. The obtained overlap matrix  for Gaussian basis functions is
\begin{equation}
O_{ij} = \left( \frac{A}{4\pi} \right)^{2N+1} \frac{1}{|M|},  \label{eq:overlap}
\end{equation}
where $A$ is the area of the 2D system, $M=(M_i +M_j)/2$ and $|M|$ is its determinant. 

The Hamiltonian in Eq.(\ref{Eq:MatrixEq}) includes three types of matrix elements, $H_{ij} = \langle \phi_i | K + U + V | \phi_j \rangle$, denoting kinetic, $U$-modified, and potential energies, respectively. The kinetic matrix element between basis states $i$ and $j$ is
\begin{equation}
K_{ij} = \left( \frac{w_0}{2m_0}  + \frac{S_w}{2 m_v}+   \sum_{\ell=1}^{N} \frac{w_{\ell} - w_{\bar{\ell}}}{2m_\ell}    \right) O_{ij}\,. \label{eq:Kij}
\end{equation}
$w_0 = W_{0,0}$, $w_\ell = W_{\ell,\ell}$, and  $w_{\bar{\ell}} = W_{\bar{\ell},\bar{\ell}}$ are diagonal elements of $W = M^{-1}$.   The kinetic energy of the photoexcited electron is linked to $w_0$, and that of the VB hole to the sum of matrix elements in $W$ ($S_w$). The kinetic energy of the CB electron-hole pair is linked to $w_{\ell} - w_{\bar{\ell}}$, representing  the electron energy above the Fermi level minus that of the missing electron below the Fermi level. 

The matrix elements for the modified band potentials in Eq.(\ref{Eq:ArtiPoten}) are given by
\begin{equation}
U_{ij} = \left(  \sum_{\ell = 1}^{N} \left( 1 - e^{-\beta_\ell} +    e^{-\beta_{\bar{\ell}}} \right)  U_0  - \frac{g_\ell w_\ell - g_{\bar{\ell}}  w_{\bar{\ell}}}{2 m_\ell}   \right)   \,\, O_{ij}. \label{eq:correction}
\end{equation}
$\beta_\ell = k_{F,\ell}^2/w_\ell $, $\beta_{\bar{\ell}} = k_{F,\ell}^2/w_{\bar{\ell}} $, $g_\ell = 1 - e^{-\beta_\ell} \left( 1+\beta_\ell \right)$ and  $g_{\bar{\ell}} =  e^{-\beta_{\bar{\ell}}} \left( 1+\beta_{\bar{\ell}} \right)$. $k_{F,\ell}$ is the Fermi wavenumber at the $\ell^{th}$ energy pocket.

The potential-energy matrix elements include two parts 
\begin{equation}
V_{ij} = \sum_{\lambda=0,\lambda< \eta }^{2N} V_{ij}^{\lambda \eta} + \sum_{\lambda=0}^{2N} V_{ij}^{\lambda }\,\,, \label{eq:Vsum}
\end{equation}
where the first term is the interaction between two quasiparticles $\{ \lambda, \eta \}$ and the second one is the interaction between quasiparticle $\lambda$ and the VB hole. The matrix element for the interaction between two quasiparticles $\{ \lambda, \eta \}$ is obtained from  
\begin{eqnarray}
V_{ij}^{\lambda \eta} &=& \sum_{{\bf q}} V({\bf q}) \sum_{{\bf X}} \phi_j\left({\bf X} \right)  \cdot \nonumber \\ &&  \phi_i^* \left({\bf X}_0, ..., {\bf X}_\lambda + {\bf q}, ... , {\bf X}_\eta - {\bf q}, ..., {\bf X}_\text{2N} \right) . \,\,\,  
\end{eqnarray}
Momentum conservation is readily seen in this Coulomb scattering process;  quasiparticle $\lambda$ is scattered from  ${\bf X}_\lambda$ to $\left( {\bf X}_\lambda +{\bf q} \right)$   whereas   quasiparticle $\eta$ is scattered from  ${\bf X}_\eta$ to $ \left( {\bf X}_\eta -{\bf q} \right)$. Calculation of the matrix elements with Gaussian basis functions yields
\begin{equation}
V_{ij}^{\lambda \eta} = O_{ij} \sum_{\bf q} V({\bf q}) e^{-\gamma_{ij}^{\lambda \eta} q^2/2}\,,
\label{Eq:PotenMa1}
\end{equation}
where $\gamma_{ij}^{\lambda \eta} = D_{\lambda \lambda} +  D_{\eta \eta} - D_{\lambda \eta} - D_{\eta \lambda} $ with $D= M_i - \frac{1}{2} M_i^\text{T} \,\, W \,\, M_i  = \frac{1}{4} \left(  M_i^\text{T} \,\, W \,\, M_j +  M_j^\text{T} \,\, W \,\, M_i  \right)$. The equation for $V_{ij}^{\lambda }$ is the same as Eq.(\ref{Eq:PotenMa1}) but with $\gamma_{ij}^{\lambda} = D_{\lambda \lambda}$ instead of $\gamma_{ij}^{\lambda \eta} $.

\subsubsection{The Keldysh-Rytova potential}
When dealing with two-dimensional (2D) semiconductors, the Keldysh-Rytova potential is a good candidate to describe the Coulomb potential \cite{Rytova_PMPA67,Keldysh_JETP79,Cudazzo_PRB11},
\begin{equation}
V_{KR}(q) = \frac{2 \pi e_\lambda e_\eta}{(1 + r_0 q)  \epsilon_b q}.
\label{Eq:KeldyshPo}
\end{equation} 
$e_\lambda$ is the charge of quasiparticle $\lambda$. The dielectric function, $\epsilon(q) = (1 + r_0 q)\epsilon_b$,  includes the dielectric constant of the surrounding barriers, $\epsilon_b$,  and polarizability of the 2D semiconductor, $r_0$. Substituting Eq.~(\ref{Eq:KeldyshPo}) in (\ref{Eq:PotenMa1}) yields
\begin{equation}
V_{ij}^{\lambda \eta} =   \frac{e_\lambda e_\eta}{2 \epsilon_b r_0} e^{-\frac{\gamma^{\lambda \eta}_{ij}}{2 r_0^2}} \left( \pi \text{Erfi} \left( \sqrt{\frac{\gamma^{\lambda \eta}_{ij}}{2 r_0^2}} \right)  - \text{Ei}\left(\frac{\gamma^{\lambda \eta}_{ij}}{2 r_0^2}\right) \right) O_{ij}, \label{eq:matrix_KRP}
\end{equation}
where $\text{Erfi}(x)$  and  $\text{Ei}(x)$ are imaginary error function and exponential integral functions, respectively.

\subsubsection{General potential, $V(\mathbf{q})$}
When dealing with Coulomb potentials of more complicated forms (e.g., due to screening of the electron gas), the matrix element in Eq.(\ref{Eq:PotenMa1}) can be obtained numerically. However, the calculation for potentials with radial symmetry can be sped  up by expanding the potential in the range  $[0,q_c]$  in the form of Fourier–Bessel series as
\begin{equation} 
V(q) = \sum_{n}^{N_\text{e} } c_n J_0 \left(b_n \frac{q}{q_c}\right) . \label{eq:bessel}
\end{equation}
$b_n$ is the $n^{th}$ root of the equation $J_0(x) = 0$ and the   coefficient $c_n$  is found from 
\begin{eqnarray}
 c_n &=&  2 \frac{ \int_0^{q_c}   q V \left(q \right)  J_0 \left( b_n \frac{q}{q_c} \right) dq }{  q_c^2  J^2_1 \left(b_n  \right)  } .         
\end{eqnarray}
The cutoff momentum $q_c$ is chosen large enough, so that the contribution from short-range scattering with larger transferred momentum can be neglected.   The number of terms in the series expansion, $N_\text{e}$, is chosen to fit the potential reasonably well, especially in the long wavelength limit ($q \rightarrow 0$).

\begin{figure}
\centering
\includegraphics*[width=9cm]{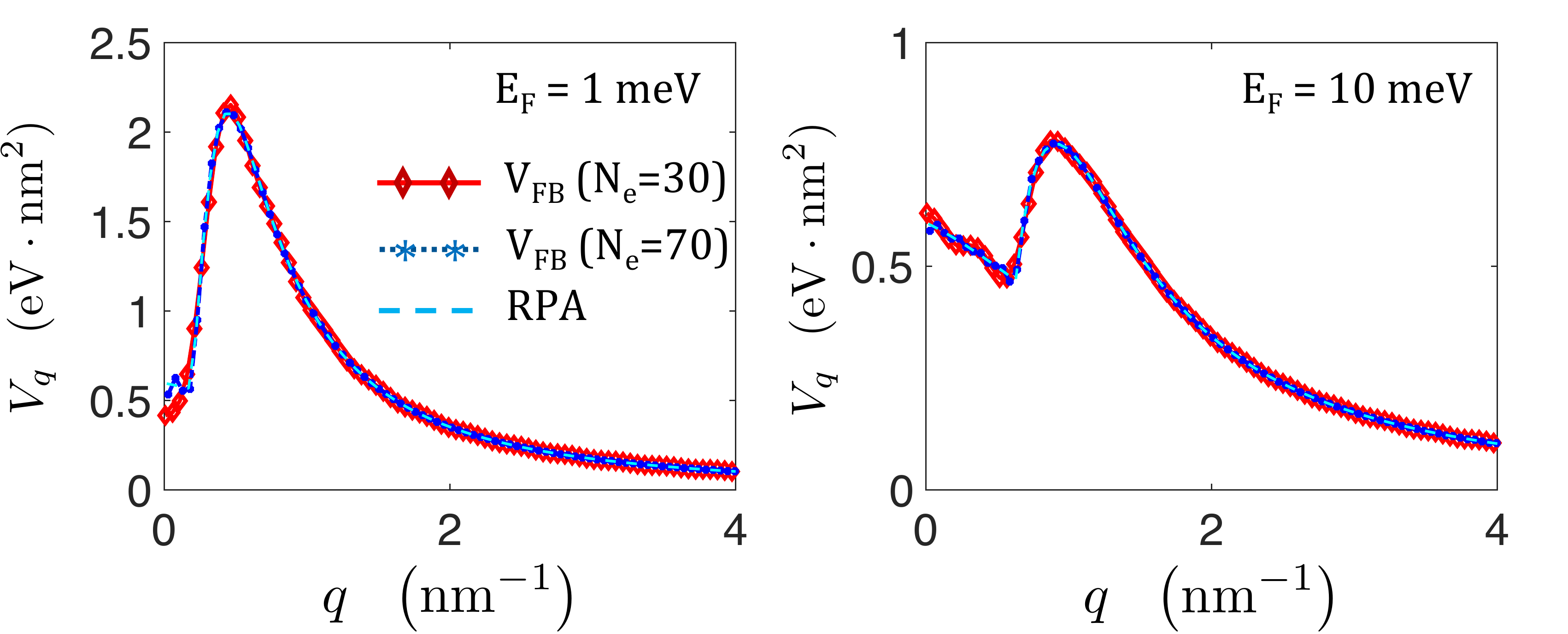}
\caption{Fourier-Bessel fits of the RPA potential when the Fermi energy is 1 and 10~meV.  The dashed cyan line is the result of Eq.~(\ref{eq:RPA}), using $\epsilon_b  =3.8$, $r_0 = 1.18$~nm, and $m_b=0.4m_0$. Lines with red diamond and blue asterisk symbols correspond to Fourier-Bessel series of the potential with 30 and 70 terms in the expansion, respectively.}\label{fig:Fig2} 
\end{figure}

Figure \ref{fig:Fig2} shows the Fourier–Bessel  fits for two different cases.  The left and right panels of the figure show results for the statically-screened potential in random phase approximation (RPA) when $E_F=1$ and 10~meV, respectively. The potential has the form
\begin{eqnarray}
V_{\text{RPA}}({\bf q}) =\frac{2\pi e^2}{A }   \frac{ 1  }{ \kappa_q + (1 + r_0 q)  \epsilon_b q  }, \label{eq:RPA}
\end{eqnarray}
where its dependence on Fermi energy comes from the Thomas-Fermi wavenumber 
\begin{eqnarray}
\kappa_q = \frac{2m_be^2}{ \hbar^2 } \left[ 1- \sqrt{1- \frac{4 k_F^2}{q^2}} \,\,\,\Theta \left(q - 2 k_F \right)\right].
\end{eqnarray}
$m_b$ is the effective mass of electrons in the Fermi sea. Figure \ref{fig:Fig2}  shows that the RPA potentials are well-fitted with series in which  $N_\text{e} <100$.

Finally, substituting Eq.~(\ref{eq:bessel}) in (\ref{Eq:PotenMa1}),  the potential matrix elements become
\begin{eqnarray}
V_{ij}^{\lambda \eta} =O_{ij}   \,\,\, \frac{1}{2\pi \gamma^{\lambda \eta}_{ij}  }  \sum_{n=1}^{N_\text{e}} c_n \exp \left(  - \frac{b^2_n}{2 \gamma^{\lambda \eta}_{ij}  q^2_\text{c}}  \right).   \label{eq:BFform}
\end{eqnarray}
The advantage of using this sum is that it requires less computation compared with numerical integration of  Eq.~(\ref{Eq:PotenMa1}). The motivation for choosing the Fourier–Bessel series over other expansion forms is the simplicity of the outcome result in Eq.~(\ref{eq:BFform}), where there is no need to invoke special functions such as the ones needed in  Eq.~(\ref{eq:matrix_KRP}).

\begin{figure}
\centering
\includegraphics*[width=6.5cm]{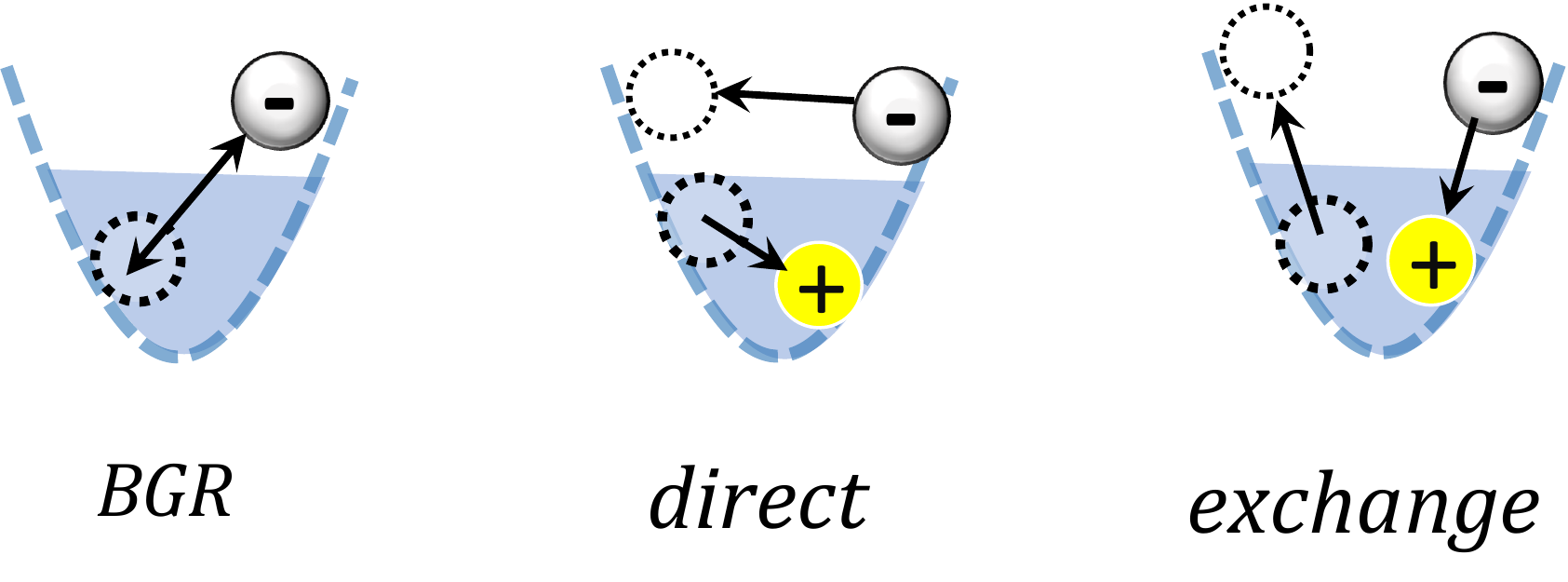}
\caption{Coulomb processes between an electron outside the Fermi sea with electrons inside the sea (left scheme), and with its CB hole (middle and right schemes). }\label{fig:exchange} 
\end{figure}

\subsection{BGR and electron-hole exchange}
So far, the effect of the Fermionic electron gas was introduced through direct coulomb interaction with CB holes. The middle scheme in Fig.~\ref{fig:exchange}  shows the direct interaction when the electron and CB hole belong to the same reservoir. The interaction is similar for other cases wherein the interacting particles reside in different valleys or having different spin configuration. The matrix elements that result from the direct interaction were analyzed  in  Sec.~\ref{sec:matrix_elements}. 

There are two more contributions we should take into account, where both stem from exchange interaction between alike electrons (i.e., similar spin and valley quantum numbers). The first consideration is the exchange interaction between electrons outside and inside the Fermi sea, as shown by the left panel of Fig.~\ref{fig:exchange}. When the electron outside the sea is bound to an excitonic complex, this  interaction helps to keep electrons from the Fermi sea away from the complex. The outcome is the celebrated BGR effect \cite{Scharf_JPCM19}, manifested in the self-energy of the electron by
\begin{equation}
\Sigma({\bf k}) = - \sum_{q} V({\bf q}) f({\bf k - q}).
\label{Eq:BGR}
\end{equation}
$f({\bf k})$ is the Fermi-Dirac distribution, and $V({\bf q})$ is the Coulomb potential. After calculating the self-energy, the matrix element due to BGR of particle $\alpha$ is
\begin{eqnarray}
L_{ij}^\alpha  &=& \sum_{\mathbf{X}} \Sigma({\bf k}_{\alpha})  \phi^\ast_i\left({\bf X} \right)  \phi_j\left({\bf X} \right)   \nonumber \\
&=&  2 \gamma_\alpha O_{ij} \int dk k \Sigma({\bf k}) \exp\left( -\gamma_\alpha k^2\right),
\label{Eq:M_BGR}
\end{eqnarray}
where $\gamma_\alpha^{-1} =\left(M^{-1}\right)_{\alpha\alpha}$ and $M = (M_i + M_j)/2$. The BGR is effective when the kinetic electron  (outside the sea) is not accompanied with a CB hole. If the latter is present, the  BGRs of the electron and CB hole offset each other (i.e., the sum from the self energies of the electron and missing electron is small). Instead of BGR, one has to consider the exchange interaction between the electron and CB hole, shown in the right panel of  Fig.~\ref{fig:exchange}.

Similar to BGR, the electron-hole exchange is only relevant when the electron and CB hole are from the same reservoir. Whereas the direct interaction  between the electron and CB hole is attractive in nature, the electron-hole exchange interaction is repulsive and weaker.  As shown by the  right panel of  Fig.~\ref{fig:exchange},  the electron-hole exchange  interaction happens when the pair recombines and excites a new  pair. Following the notation of Eq.(\ref{Eq:TrialWave-k}), the initial and final pairs correspond, respectively, to  $c^\dagger_{{\bf k}_\ell}  c_{{\bf p}_\ell} |  \varphi_0 \rangle$ and $c^\dagger_{{\bf k}'_\ell}  c_{{\bf p}'_\ell} |  \varphi_0 \rangle$. The magnitude of this interaction is proportional to $V({\bf k}_\ell - {\bf p}_\ell )$ because the electron jumps from state  ${\bf k}_\ell$ to fill  the empty (hole) state $  {\bf p}_\ell $ inside the Fermi sea.  Translation symmetry mandates momentum conservation, and thus,  ${\bf k}_\ell - {\bf p}_\ell = {\bf k}'_\ell - {\bf p}'_\ell$.  The resulting matrix element is calculated using the transformation $ \{ {\bf k}_\ell , {\bf p}_\ell\}  \rightarrow \{ {\bf k}_\ell, {\bf p}^*_\ell =  {\bf k}_\ell - {\bf p}_\ell\} $. In turn, the coordinates change as ${\bf X} \rightarrow {\bf Z} = T {\bf X}$  where all matrix elements of  $T$ are zeros except $T_{\lambda,\lambda} = 1$ for $\lambda \neq \bar{\ell}$, $T_{\bar{\ell},\bar{\ell}} = -1$ and  $T_{\bar{\ell},\ell} = 1$. In addition, the basis functions change  accroding to
\begin{equation}
\phi_i({\bf X}) \rightarrow  \phi'_i({\bf Z}) = \exp \left( -\frac{1}{2} {{\bf Z}}^{\text{T}} \, B^i  \, {\bf Z}\right),
\end{equation} 
where $B^i = \left(T^{-1} \right)^\text{T} M_i T^{-1}$. The exchange energy coming from the pair $\{\ell,\bar{\ell}\}$ is given by 
\begin{equation}
\Xi_{ij}^{\ell,\bar{\ell}} = \sum_{{\bf Z}'} V({\bf Z}_{\bar{\ell}}) \left(   \sum_{{\bf Z}_{\ell }}  \phi'_i({\bf Z}) \right) \left(   \sum_{{\bf Z}_{\ell }}  \phi'_j({\bf Z}) \right),
\end{equation}
where the summation ${\bf Z}'$ means a sum of all its components but ${\bf Z}_\ell$. Next, we define the matrix $F^i$, 
\begin{equation}
F^i_{\lambda\eta} = B^i_{\lambda\eta} - \frac{B^i_{\lambda\ell} \,\,  B^i_{\ell \eta} }{B^i_{\ell\ell}}, 
\end{equation}
whose elements in the $\ell^\text{th}$ row and column are zero.  Getting rid of these row and column we get a new $(2N)\times (2N)$ matrix $G^{i}$. 
The matrix element for the electron-hole exchange interaction  is then given by
\begin{eqnarray}
\Xi_{ij}^{\ell,\bar{\ell}} &=& \left( \frac{A}{4\pi} \right)^{2N+1} \!\!\!\!\!\! \frac{4}{B^i_{\ell\ell} B^j_{\ell\ell}} \frac{\gamma_{ij}}{\det G}   \sum_{{\bf Z}_{\bar{\ell}}}  V({{\bf Z}}_{\bar{\ell}})  e^{-\gamma_{ij} {\bf Z}^2_{\bar{\ell}}}\,.\,\,\,\,\, \label{eq:ehx}
\end{eqnarray}
$G = \frac{1}{2} ( G^{i} +G^{j})$ and $1/\gamma_{ij} = G^{-1}_{\bar{\ell} \bar{\ell}}$. When using the Keldysh-Rytova potential  in Eq.(\ref{Eq:KeldyshPo}), the electron-hole exchange matrix element becomes
\begin{eqnarray}
\Xi_{ij}^{\ell,\bar{\ell}} &=& \left( \frac{A}{4\pi} \right)^{2N+1} \frac{4}{B^i_{\ell\ell} B^j_{\ell\ell}}\, \frac{\gamma_{ij}}{| G|} \, \frac{2e^2}{\epsilon_v r_0} e^{-\frac{\gamma_{ij}}{r_0^2}} \nonumber \\
 &\times&   \left( \pi \,\, \text{Erfi}\left(\sqrt{\frac{\gamma_{ij}}{r_0^2}}\right)-\text{Ei}\left(\frac{\gamma_{ij}}{r_0^2}\right)\right).
\end{eqnarray}

\subsection{Computation details}

To find the ground state of the system, we solve Eq.~(\ref{Eq:MatrixEq}) with the help of Eqs.~(\ref{eq:overlap})-(\ref{eq:Vsum}) for the overlap, kinetic, modified potential, and Coulomb potential matrix elements.  The Coulomb matrix elements in Eq.~(\ref{eq:Vsum}) are calculated from Eq.~(\ref{eq:matrix_KRP}) or (\ref{eq:BFform}) when using the Keldysh-Rytova potential or a general potential form, respectively. The BGR matrix element in Eq.~(\ref{Eq:M_BGR}) is added to electrons that are not accompanied with CB holes, and the electron-hole exchange matrix element in Eq.~(\ref{eq:ehx}) is added to each electron-hole CB pair.  

The computation time needed to  find the ground state of the system from Eq.~(\ref{Eq:MatrixEq}) depends on the efficiency of the search process for variational parameters \cite{Varga1997,Varga2008,VargaBook}.  Here, the variational parameters are all  elements of the matrices  $M_i$ in the Gaussian basis functions. We briefly discuss the trisection method which is example for an oriented search process and then provide the needed steps to perform the overall calculation.

\begin{figure}
\centering
\includegraphics*[width=8cm]{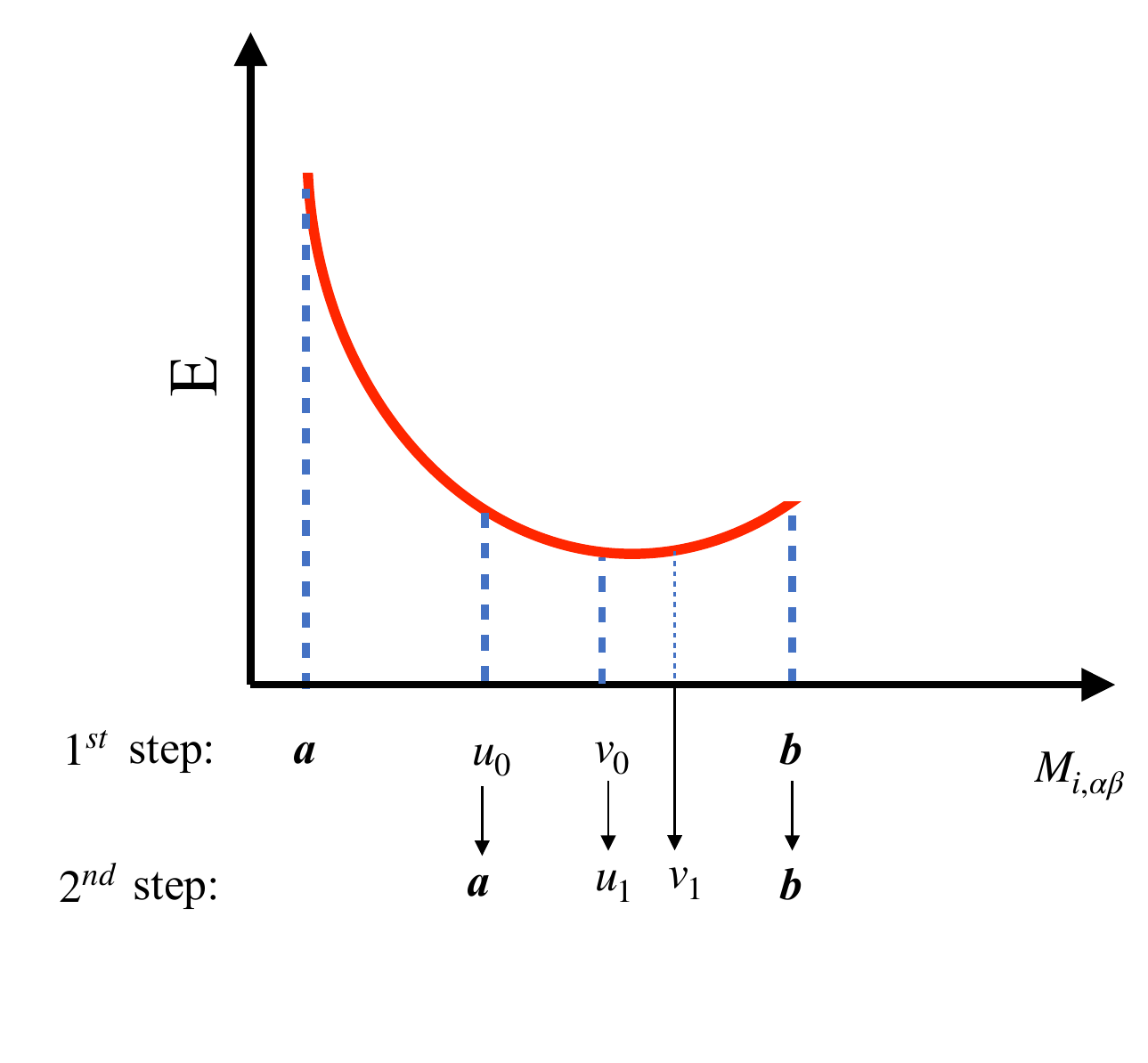}
\caption{ Trisection method  to obtain the optimal value for the variational parameter $M_{i,\alpha\beta}$ }\label{fig:TriSec} 
\end{figure}

\subsubsection{Trisection Method}
 
Figure~\ref{fig:TriSec} shows the trisection method to find the  minimum  energy as a function of the matrix element, $E(M_{i,\alpha\beta})$, in an interval $[a,b]$. We choose two points $u<v$ in this interval and calculate $\{E(u)$,  $E(v)\}$. If $E(u)<E(v)$ then the minimum is in $[a,v]$.  Otherwise, the minimum is in $ [u,b]$. To save computational effort,  $u$ and $v$ are chosen by using the data from the calculated values. We define $u=a+\rho (b-a)$ and $v=b-\rho (b-a)$ where $0<\rho<1$ is a constant.  The model is  simplified by defining  the  length unit as  $(b-a)$, such that $a=0, b=1$. The first step starts with   $\{u_0=\rho $, $v_0=1-\rho \}$. In the next step we have $u_1 \equiv v_0$ if $E(u_0)>E(v_0)$, as shown in Fig.~\ref{fig:TriSec}, or    $ v_1 \equiv u_0 $ if $E(u_0)<E(v_0)$. Both lead to the equation $\rho=\left( 1- \rho\right)^2 $ whose solution is
 \begin{equation}
 \rho^* = \frac{3-\sqrt{5}}{2} \simeq 0.382.
\end{equation}

The process to find an optimal value for each $M_{i,\alpha\beta}$ is as follows.
\begin{enumerate}

\item Start with an initial range $[a_0,b_0]$ in which we expect to find an optimal value for $M_{i,\alpha\beta}$. Calculate $\{E(u_0)$, $E(v_0)\}$ for   $\{u_0=a_0+\rho^* (b_0-a_0)$, $v_0=b_0-\rho^* (b_0-a_0)\}$.

\item Compare and choose values for the next step:   If $E(u_0) < E(v_0)$ then $\{ a_1=a_0, b_1=v_0, u_1=a_1+\rho^* (b_1-a_1), v_1=u_0 \}$, otherwise  $\{ a_1=u_0, b_1=b_0, u_1= v_0, v_1= b_1-\rho^* (b_1-a_1) \}$.

 \item Repeat step 2 until the convergence condition, $b_n \!-\! a_n \!< \!\epsilon $, is met for a chosen resolution $\epsilon$. 
\end{enumerate}

\subsubsection{Adding basis functions}
 To continue lowering the energy of the system,  basis functions are added to the existing set as follows. Assuming we already have $k -1$ basis functions, where each corresponds to a matrix $M_i$ ($ i=1,2,..., k-1$), the steps to add the ${k}^\text{th}$ basis function are:
\begin{enumerate}
\item Randomly generate the matrix elements of $M_{k}$. 
\item The matrix elements $M_{k, \alpha \beta}$ are optimized using the trisection method to minimize the energy function $E(M_{k,\alpha\beta})$ using Eq.(\ref{Eq:MatrixEq}). 
\item The previous step is repeated several times to guarantee optimal selection of the matrix   $M_{k}$, after which it is added to the basis set. 
\end{enumerate}

Once the basis set comprises (a pre-determined) ${\mathcal{N}_{\text{b}}}$ matrices, the trisection method is re-employed to further minimize the energy function $E(M_{k,\alpha\beta})$ using Eq.(\ref{Eq:MatrixEq}) for each element of each matrix in the set ($1 \leq k \leq \mathcal{N}_{\text{b}}$).  This process can be repeated several times until the improvement is marginal.

\subsection{Kinetic energies, average distances, and density distributions}
After obtaining the wave function of the system, we can extract individual quasiparticle properties. First, we define the normalization factor,
\begin{equation}
\mathcal{N} = \sum_{i,j} C_i^* C_j  O_{ij},
\end{equation}
where the sums over $i$ and $j$ run over the basis functions ($1 \leq  i,j \leq \mathcal{N}_\text{b}$). The  charge distribution of particle $\alpha$ in momentum space is
\begin{equation}
\rho_{\alpha}(k) =  \frac{4\pi}{A\mathcal{N}} \sum_{i,j} C_i^* C_j  O_{ij} \cdot  \gamma_{ij}^\alpha e^{- \gamma_{ij}^\alpha k^2}.
\end{equation}
$\gamma^\alpha_{ij} = 1/(W_{ij})_{\alpha \alpha}$, where the matrix $W_{ij}$ is the inverse of $M_{ij}=(M_i+M_j)/2$. When $\alpha$ is the VB hole, we get a similar result but with $\gamma^\alpha_{ij} = 1/(T_{ij})_{\alpha \alpha}$, where the matrix $T_{ij}$ is the inverse of $P^TM_{ij}P$ and 
 \begin{equation}
 P=
\begin{pmatrix} 
-1 & -1 & -1 & ...  &  -1  \\
1  & 0 &  0  & ... & 0  \\
0  & 1 &  0  & ... & 0  \\
\vdots &\vdots  & \vdots  & \vdots   \\
 0 & 0 & ... & 1 & 0  \\
\end{pmatrix} .
\end{equation}

Using Eq.~(\ref{eq:Kij}), the kinetic energy of quasiparticle $\alpha$ is  
\begin{equation}
K^\alpha =\frac{1}{\mathcal{N}} \sum_{i,j} C_i^* C_j \frac{w_{\alpha} }{2 m_\alpha} O_{ij}. 
\end{equation}
Similarly,  the kinetic energy of the VB hole is
\begin{equation}
K^v = \frac{1}{\mathcal{N}}\sum_{i,j} C_i^* C_j \frac{S_w}{2 m_v}  O_{ij}. 
\end{equation}

\subsubsection{Relative distances}
To calculate average distances between quasiparticle $\alpha$ and $\beta$, we write the basis function in real space $\phi_i({\bf x}) = \phi_i({\bf r}_0,{\bf r}_1, ..., {\bf r}_{2N+1})$,
\begin{eqnarray}
\phi_i (\mathbf{x}) &=&  \frac{e^{i {\bf Q} \cdot {\bf r}_v }}{  A^{N+1}}  \sum_{{\bf X}} \phi_i({\bf X})e^{i \sum_{\alpha}{\bf X}_\alpha \cdot ({\bf r }_\alpha - {\bf r }_v)},  
\label{Eq:kSpaceDis1}
\end{eqnarray}
where ${\bf r}_v$ is the position vector of the VB hole and ${\bf X}_\alpha$ is the momentum vector of quasiparticle $\alpha$. Substituting Eq.~(\ref{Eq:GaussBas}) in (\ref{Eq:kSpaceDis1}), we obtain
\begin{equation} 
 \phi_i ( {\bf x}) = \left( \frac{\sqrt{A}}{2\pi} \right)^{\text{2N+1}}  \!\!\!\!\!\!\! \frac{1}{ \sqrt{A}\,\, |M_i|  }   \exp \left( -  \frac{1}{2} {\bf x}^\text{T} \left(M_i\right)^{-1} {\bf x}\right),     
 \label{Eq:RealWave}  
\end{equation}
where ${\bf x}^\text{T} = \{ {\bf r}_1  - {\bf r}_v, ... ,  {\bf r}_{2N+1}  - {\bf r}_v \} $. Using real-space basis functions, the average distance between quasiparticles $\alpha$ and $\beta$ becomes
\begin{equation} 
\langle r_{\alpha\beta}^2 \rangle = \langle \psi |  \left( {\bf r}_\alpha  - {\bf r}_\beta \right)^2 | \psi \rangle  = \frac{1}{\mathcal{N}} \sum_{i,j} C^*_i C_j R^{\alpha \beta}_{ij}, 
\end{equation} 
where 
\begin{eqnarray}
R^{\alpha \beta}_{ij}= 2 O_{ij}  
 \left( \left(  M^1_{ij} \right)^{-1}_{\alpha \alpha}        +    \left(  M^1_{ij}\right)^{-1}_{\beta \beta}  - 2\left(  M^1_{ij} \right)^{-1}_{\alpha \beta}               \right)  .
\end{eqnarray}
$M^1_{ij} =  M_i^{-1} +  M_j^{-1}$. Similarly, the average distance between  the VB hole and quasiparticle $\alpha$ becomes
\begin{equation} 
\langle r_{\alpha v}^2 \rangle= \frac{1}{\mathcal{N}} \sum_{i,j} C^*_i C_j R^{\alpha v}_{ij} ,
\end{equation} 
where
\begin{eqnarray}
R^{\alpha v}_{ij}= 2 O_{ij}  
  \left( M^1_{ij} \right)^{-1}_{\alpha \alpha}        .
\end{eqnarray}

\subsubsection{Density distributions}
To calculate the density distribution $\rho({\bf r}_{\alpha \beta})$ for the separation  ${\bf r}_{\alpha \beta}$ of quasiparticles $\alpha$ and $\beta$, we need to transform the wave function in Eq.(\ref{Eq:RealWave}) from $ {\bf x}= \{ {\bf r}_1  - {\bf r}_v, ... ,  {\bf r}_{2N+1}  - {\bf r}_v \}^\text{T} $  to ${\bf y} = \{ {\bf r}_1  - {\bf r}_\beta, ... ,  {\bf r}_{2N+1}  - {\bf r}_\beta, {\bf r}_{2N+2}  - {\bf r}_\beta\}^\text{T}   $ 
\begin{equation}
{\bf x} = U^\beta {\bf y},
\end{equation}
where  the matrix of transformation is
\begin{equation}
 U^\beta=
\begin{pmatrix} 
1 & 0 & \cdots & 0_{1,\beta}  & \cdots &  -1  \\
0  & 1 & \cdots  & 0_{2,\beta} & \cdots & -1  \\
\vdots &\vdots  & \vdots  & \vdots & \cdots   & \vdots  \\
 0_{\beta, 1}  &   0_{\beta, 2} & \cdots  &   0_{\beta, \beta} & \cdots &   -1_{\beta, 2N+2}  \\
\vdots &\vdots  & \vdots  & \vdots  & \cdots  & \vdots  \\

0 & 0 & \cdots & 0  & 1  &-1 \\
\end{pmatrix} .
\end{equation}
The basis function becomes
\begin{equation}  
 \phi^\beta ({\bf y})  = \left( \frac{\sqrt{A}}{2\pi} \right)^{\text{2N+1}}  \!\!\!\!\!\!\!\! \frac{1}{ \sqrt{A}\,\, |M_i|  }  \exp \left( -  \frac{1}{2} {\bf y}^\text{T} \,\,\, M_i^\beta \,\,\,  {\bf y}\right),
\end{equation}
where $M_i^\beta= (U^\beta)^\text{T} \left(M_i\right)^{-1} U^\beta $. Using these basis functions, the density distribution for  ${\bf y}_\alpha = {\bf r}_\alpha  -  {\bf r}_\beta$ becomes
\begin{equation}  
\rho_\beta({\bf y}_\alpha) =    \frac{1}{\pi \mathcal{N} } \sum_{i,j} c^*_i c_j \,\,\,  O_{ij} \,\,\, \Gamma_{ij}^{\alpha } \,\,\,  e^{  - \Gamma_{ij}^{\alpha} y_\alpha^2} \,\,,    
\end{equation}
where $\Gamma_{ij}^\alpha =  \left( M_{ij}^{-1} \right)_{\alpha \alpha}^{-1}$ with  $M_{ij} = \left(M_i^\beta + M_j^\beta\right)/2$.  

\section{conclusions and Outlook} \label{sec:DisCon}
We have described the SVM-$k$ model to study composite excitonic states in doped semiconductors. Many-body interactions between the electron gas and the excitonic state are manifested by introducing conduction-band holes, created when electrons are pulled out of the Fermi sea to bind the photoexcited electron-hole pair.

The number of conduction-band electron-hole pairs in the excitonic state of a given material is determined by two factors. The first one is intrinsic and dictated by the number of low-energy pockets at the edge of the conduction band in electron doped semiconductors. For example, the number of pairs can be as large as 12 in electron-doped diamond or silicon owing to their six spin-degenerate valleys (around the $X$-point in diamond or along the $\Delta$-axis in silicon). Similarly, the number of pairs can be as large as 8 in germanium owing to its  four spin-degenerate valleys around the $L$ point. Yet, more commonly observed in experiment are smaller complexes, such as in GaAs whose correlated states can host at most  two electrons owing to the nearly spin-degenerate valleys around the $\Gamma$ point. A similar behavior is observed in MoSe$_2$ monolayer owing to its spin-polarized valleys in the $K$ point and its time-reversed $-K$ point. Electron-doped WSe$_2$ monolayer  has an extra unique feature in that quantum numbers of the photoexcited electron are different than those of electrons in the Fermi sea, thereby allowing for composites with more than two electrons \cite{h}.  Excitonic states with $2(N+1)$ quasiparticles where $N>1$ can be realized if the energy difference between a composite with $n$ or $n+1$ electrons is large enough to establish robust correlations with the Fermi sea. The value of $n$ can approach the intrinsic limit $N$ by enhancing the Coulomb interaction through engineering of low dielectric-constant environments with reduced dimensionality. 

The SVM-$k$ model is general and can be used to  study various multi-valley semiconductors, either electron or hole doped. In the latter case, one has to replace the discussion of holes inside an electron Fermi sea with that of electrons inside a hole Fermi sea. Along with this work, we present three works in which results of the SVM-$k$ model are shown for composite excitonic states in electron-doped monolayer transition-metal dichalcogenides \cite{s,g,h}. The theory can be extended to study similar phenomena in graphene and  semimetals, or to study problems wherein the mobile impurity is not necessarily a photoexcited valence-band hole in electron-rich environment (or vice versa).  

\acknowledgments{This work was supported by the Department of Energy, Basic Energy Sciences, Division of Materials Sciences and Engineering under Award DE-SC0014349 (DVT), and by the Office of Naval Research under Award N000142112448 (HD).}

%

\end{document}